\documentclass[11pt,twocolumn,twoside]{pnas-new}
\setboolean{displaywatermark}{false}

\title{\bf Enhancements in cloud condensation nuclei concentrations from turbulent fluctuations in supersaturation}
\date{\vspace{-5ex}}

\author[a]{Jesse C. Anderson}
\author[b]{Payton Beeler} 
\author[c]{Mikhail Ovchinnikov}
\author[a]{Will Cantrell}
\author[d]{Steven Krueger}
\author[a]{Raymond A. Shaw}
\author[e]{Fan Yang}
\author[c,*]{Laura Fierce}

\affil[a]{Michigan Technological University}
\affil[b]{Washington University in St. Louis}
\affil[c]{Pacific Northwest National Laboratory}
\affil[d]{University of Utah}
\affil[e]{Brookhaven National Laboratory}

\leadauthor{Anderson} 

\significancestatement{Increases in cloud droplet number concentrations from human emissions of aerosol particles lead to clouds that are, on average, more reflective and longer lived. Large Eddy Simulations and Earth System Models are used to quantify these aerosol-cloud interactions, but the spatial and temporal resolution of these models are too coarse to represent the impact of turbulence at the smallest scales. We show that small-scale turbulent fluctuations lead to cloud droplet formation even when air is, on average, subsaturated, which would be impossible in conventional models of cloud microphysics. Our findings suggest that models that neglect turbulent fluctuations in supersaturation will underestimate cloud condensation nuclei concentrations under specific supersaturation regimes, leading to error in predicted aerosol-cloud interactions.}  
\authorcontributions{L.F. and J.C.A. designed research. J.C.A., L.F., P.B., M.O, and and F.Y. performed research. J.C.A., L.F., W.C., S.K, and R.A.S. analyzed data. J.C.A. and L.F. wrote the manuscript. L.F., J.C.A., P.B., M.O., W.C., S.K, R.A.S, and F.Y. reviewed and edited the manuscript.}
\authordeclaration{The authors declare no competing interest.}
\correspondingauthor{\textsuperscript{2}To whom correspondence should be addressed. E-mail: laura.fierce@pnnl.gov}

\keywords{turbulence $|$ CCN activation $|$ cloud aerosol interaction $|$ supersaturation}

\begin{abstract}
    Changes in the properties and distribution of clouds from emissions of aerosol particles is a large source of uncertainty in predictions of weather and climate. These aerosol-cloud interactions depend critically on the ability of aerosol particles to activate into cloud condensation nuclei (CCN). A key challenge in modeling CCN activation and the formation of cloud droplets is the representation of interactions between turbulence and cloud microphysics. Turbulent mixing leads to small-scale fluctuations of water vapor and temperature that are not resolved in large-scale atmospheric models. We used Lagrangian parcel simulations driven by a high-resolution Large Eddy Simulation of a convective cloud chamber to quantify the impact of these small-scale fluctuations on CCN activation. We show that small-scale fluctuations in environmental properties strongly enhance CCN activation, which suggests that conventional Large Eddy Simulations and Earth System Models that neglect these fluctuations underestimate cloud droplet formation.
\end{abstract}

\dates{This manuscript was compiled on \today}
\doi{\url{www.pnas.org/cgi/doi/10.1073/pnas.XXXXXXXXXX}}

\begin{document}

\ifthenelse{\boolean{shortarticle}}{\ifthenelse{\boolean{singlecolumn}}{\abscontentformatted}{\abscontent}}{}


\twocolumn[
\maketitle
  \begin{@twocolumnfalse}
The effect of aerosol emissions on the properties and distribution of clouds is a large source of uncertainty in predictions of weather and climate. These aerosol-cloud interactions depend critically on the ability of aerosol particles to activate into cloud condensation nuclei (CCN). A key challenge in modeling CCN activation and the formation of cloud droplets is the representation of interactions between turbulence and cloud microphysics. Turbulent mixing leads to small-scale fluctuations of water vapor and temperature that are not resolved in large-scale atmospheric models. We used Lagrangian parcel simulations driven by a high-resolution Large Eddy Simulation of a convective cloud chamber to quantify the impact of these small-scale fluctuations on CCN activation. We show that small-scale fluctuations in environmental properties strongly enhance CCN activation, which suggests that conventional Large Eddy Simulations and Earth System Models that neglect these fluctuations underestimate cloud droplet formation.
\vspace{12pt}
 \end{@twocolumnfalse}]

\footnotetext[1]{To whom correspondence should be addressed. E-mail: laura.fierce@pnnl.gov}

\section*{Introduction}
Aerosol-cloud interactions are the largest source of uncertainty in predictions of radiative forcing relative to the pre-industrial atmosphere. Increases in cloud droplet number concentrations from human emissions of aerosol particles lead to clouds that are, on average, more reflective and longer lived \cite{twomey1977influence,albrecht1989aerosols,rosenfeld2008flood}. These processes are partially determined by how efficiently aerosol particles activate into cloud condensation nuclei (CCN). At a fixed supersaturation or in a uniform updraft, activation is well modeled by $\kappa$-K\"{o}hler theory \cite{petters2007single}. However, CCN activation in turbulent conditions is not well understood. 

Clouds are turbulent on length scales from $\sim$1~mm to beyond the size of the cloud. A key characteristic of turbulence is the chaotic motion of the air, accelerating the mixing of air masses. Mixing of air parcels within a cloud causes fluctuations in the water vapor mixing ratio and temperature, leading to fluctuations in water vapor supersaturation, $s$ \cite{Gerber1991,Kulmala1997,korolev2000drop,Seibert2017, anderson2021effects}. Despite a number of studies showing that fluctuations in $s$ impact cloud droplet formation and the cloud droplet size distribution \cite{Chandrakar2016,siewert2017statistical,chen2018bridging,saito2019broadening,macmillan2022direct}, these fluctuations are often neglected in large-scale atmospheric models. 

Earth System Models represent the globe using a collection of grid boxes that are on the order of 100~km in each horizontal dimension. Even Large Eddy Simulation (LES) models that are designed to study cloud systems in detail use grid spacing on the order of 10--100~m, which is still too coarse to resolve the small-scale fluctuations in temperature and water vapor induced by turbulence within clouds. While LES models are used to parameterize subgrid-scale (SGS) cloud processes \citep{golaz2002pdf1, golaz2002pdf2, bogenschutz2013simplified}, these parameterizations represent spatial variability on the scale of the LES grid, not variability resulting from small-scale turbulence. While models that represent SGS variability in environmental properties on cloud microphysics have been developed \cite{grabowski2017broadening, abade2018broadening,hoffmann2019inhomogeneous, chandrakar2022supersaturation}, the impact of small-scale fluctuations in supersaturation on CCN activation spectra has not been well quantified.

To explore the impact of unresolved SGS fluctuations on CCN activation within warm, non-precipitating clouds, we combine Lagrangian particle tracking with a high-resolution LES of the Michigan Technological University (MTU) Pi Chamber \citep{Thomas2019,yang2022large}. The Pi Chamber is a convection cloud chamber, with a volume of $\pi$~m$^3$, which is designed to study CCN activation, cloud droplet growth, and turbulence-induced variability in environmental properties \cite{chang2016laboratory}. The LES of the Pi Chamber is a modified version of the SAM model \cite{khairoutdinov2003cloud}, with a grid spacing of 3.125 cm in each dimension. This small grid spacing is several orders of magnitude smaller than cloud-scale LES, resolving fluctuations in environmental properties at very small scales.

\section*{Results}
We used the high-resolution LES to drive Lagrangian trajectories of air parcels within the Pi Chamber. Velocity fields from the LES were used to propagate parcels throughout the chamber, while treating the particles as passive tracers. Gravitational settling is modeled by a first-order decay in particle number concentrations, which depends on their terminal settling velocity. 

\begin{figure}[ht] 
 \begin{center}
 \includegraphics[width=3.5in]{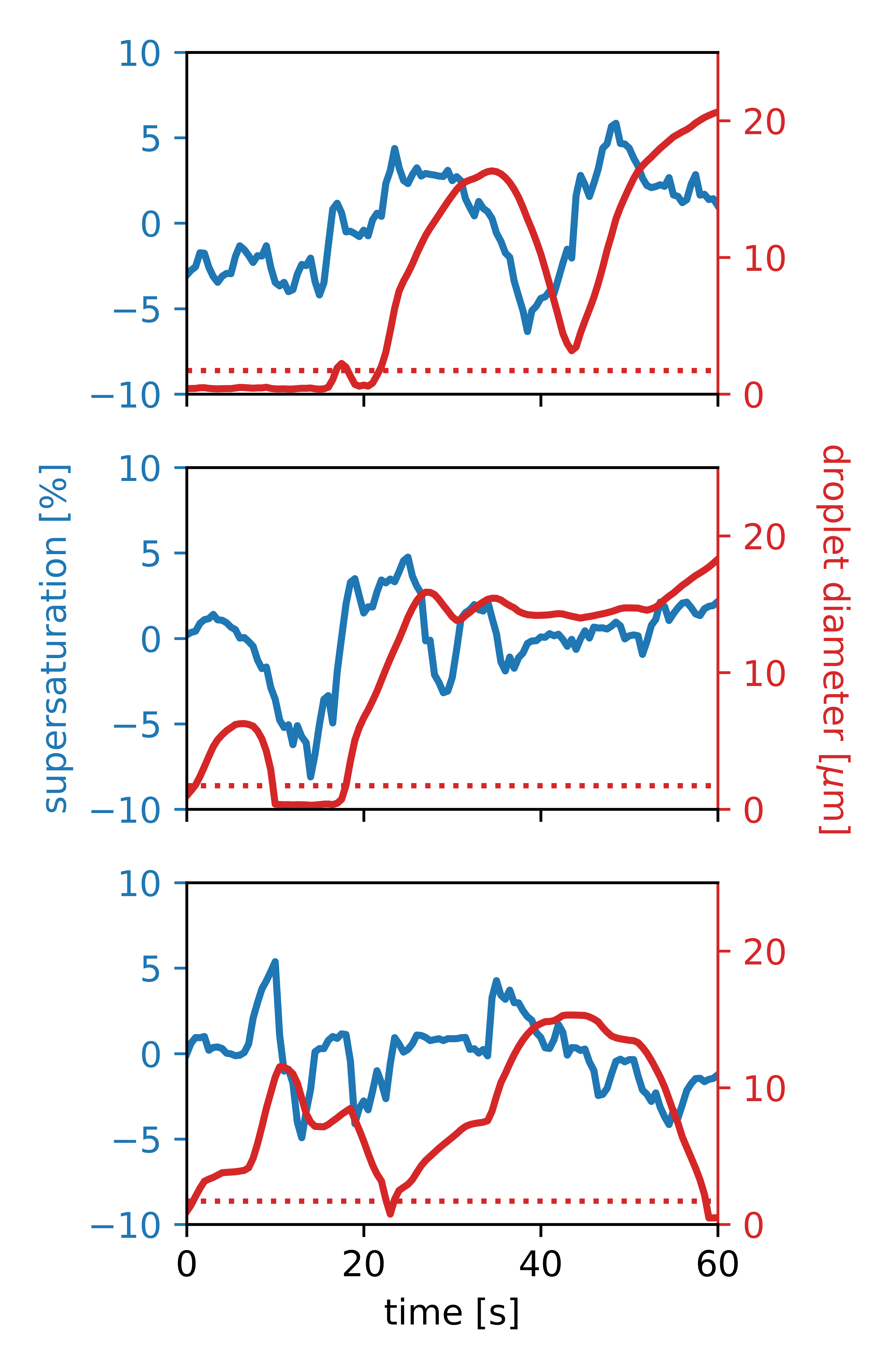}
 \end{center}
  \caption{The temporal evolution of the supersaturation (blue), and the mean droplet diameter (red) along three Lagrangian parcels in the Pi Chamber. The dashed red lines show the critical diameter for CCN activation. In each panel, the particle activates and grows in response to the supersaturation fluctuations. In this case, $\overline{s}$=-2.7\% in the chamber domain, which means that no particles would be CCN-active if $s$ was uniform.
  }
  \label{fig:time_series_S_Dp}
\end{figure}

We simulated 10,000 Lagrangian trajectories within the Pi Chamber and, along each trajectory, extracted the temperature and water vapor mixing ratio from the LES fields. The temporal evolution of temperature and water vapor experienced by each parcel was used to compute $s(t)$, which is shown for three example particles by blue lines in Fig.~\ref{fig:time_series_S_Dp}. The temporal evolution of the environmental properties is then used to drive box model simulations of CCN activation and growth, shown for the three example particles by red lines in Fig.~\ref{fig:time_series_S_Dp}. All particles in the simulation and in the chamber are comprised of pure NaCl and have a dry diameter of 130~nm. In this study, the environment simulated by the LES was used to drive particle growth in the box model without direct feedback on the environment. Instead, this feedback was captured by the microphysics scheme of the LES.

\subsection*{Droplet Growth Driven by Supersaturation Fluctuations}
As particles move through the chamber, they each experience a different time series of supersaturation fluctuations. The particles grow by condensation in response to the local supersaturation. To become CCN active, a particle must enter in a region where $s>s_{\text{crit}}$, where $s_{\text{crit}}$ is the particle's critical supersaturation. If a particle remains at this elevated supersaturation for enough time, the particle will grow through condensation and eventually become large enough that its diameter exceeds  its critical diameter ($D_{\text{p,crit}}$). When this condition is met, the particle is considered CCN-active and will grow without bound until the supersaturation drops to $s=0$. The critical diameter of a particle is shown by the dashed red lines in Fig.~\ref{fig:time_series_S_Dp}.  

The three examples of the particle diameter (red) and supersaturation (blue) shown in Fig.~\ref{fig:time_series_S_Dp} show how the fluctuations in $s$ lead to activation and deactivation of aerosol particles as they are transported through the chamber, despite $\overline{s}$=-2.7\% in the LES domain. For example, at a simulation time of $\approx$20~seconds, the particle shown in the top panel enters a region where $s>s_{\text{crit}}$ and remains there for enough time that it becomes CCN-active, indicated by $D>D_{\text{crit}}$. After activating, the particle enters a region that is subsaturated with water vapor ($s<0$) at $\approx$35--45~s and begins to evaporate. In this case, the droplet returns to a region were $s>0$ at a simulation time of $\approx$45~s, which occurs before the particle has time to deactivate. The particle resumes its growth by condensation.

On the other hand, the middle and bottom panels of Fig.~\ref{fig:time_series_S_Dp} show examples of particles that activate and then deactivate over the course of the simulation. In both cases, the particles are exposed to $s>s_{\text{crit}}$ near the beginning of the simulation and quickly activate. However, these particles are then exposed to air with $s<0$ shortly after the start of the simulation, leading to deactivation at $\approx$10~s and $\approx$25~s in the middle and bottom panels, respectively. The fluctuations in $s$ cause the particles to activate and deactivate over the course of the simulation.

\subsection*{CCN Activation a Turbulent Environment}
\begin{figure}[ht]
 \begin{center}
 \includegraphics[width=3.5in]{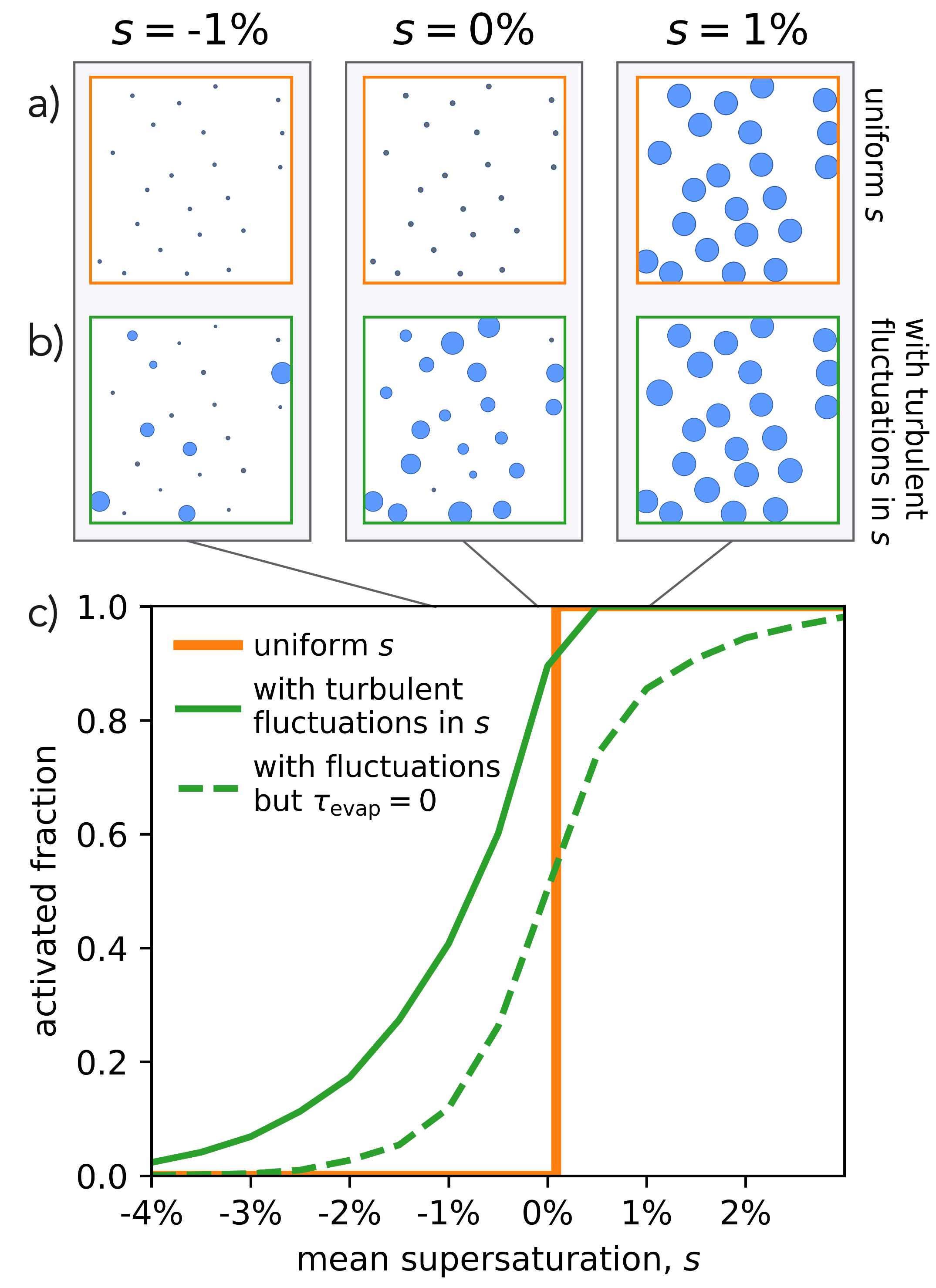}
 \end{center}
  \caption{(a) Under the uniform $s$ assumption, all particles remain CCN-inactive until $\overline{s}$ exceeds their critical supersaturation, whereas (b) a portion of particles will activate even when $\overline{s}<0\%$ when turbulent fluctuations in $s$ are included. As a result, (c) across $\overline{s}$, the fraction of particles activating tends to be greater when turbulent fluctuations are included (solid green line) than under the uniform $s$ approximation (orange line). A greater fraction of particles will activate under the dynamic simulation of turbulence (solid green line) than it would be if $\tau_{\text{evap}}=0$ (dashed green line); the solid green line shows the fraction of droplets with $D_{\text{p}}\geq D_{\text{p,crit}}$, whereas the dashed green line shows the fraction of particles with $s\geq s_{\text{crit}}$. }
  \label{fig:activated_fraction_mean_S}
\end{figure}

To explore the impact of turbulent fluctuations in $s$ on CCN activation, we repeated the Lagrangian droplet simulations at varying levels of mean environmental supersaturation ($\overline{s}$) in the LES domain. Over the range of $\overline{s}$, we calculate the fraction of particles that are CCN-activate ($D>D_{\text{crit}}$) at a snapshot in time, after 60~s of simulation. If we assume that $s$ is uniform across the LES domain, the population of identical particles takes up water uniformly as $\overline{s}$ increases and, given enough time, all activate once $\overline{s}$ exceeds $s_{\text{crit}}$, shown schematically in Fig.~\ref{fig:activated_fraction_mean_S}a. On the other hand, if turbulent fluctuations in $s$ are represented, a portion of particles will become CCN-active even if $\overline{s}$ is below $s_{\text{crit}}$, shown in Fig.~\ref{fig:activated_fraction_mean_S}b.

The fraction of particles that are CCN-active with and without turbulent fluctuations in $s$ are shown as a function of $\overline{s}$ by the solid green and orange lines in Fig.~\ref{fig:activated_fraction_mean_S}c, respectively. The fraction activated under uniform $s$ is represented by a step function with all particles activating once $\overline{s}>s_{\text{crit}}$, shown by the orange line in Fig.~\ref{fig:activated_fraction_mean_S}. On the other hand, when turbulent fluctuations in $s$ are included (solid green line in Fig.~\ref{fig:activated_fraction_mean_S}c), a large fraction of particles become CCN-active even when $\overline{s}<0$, which would be impossible under the uniform $s$ approximation. While the activated fraction when $s$ is uniform is slightly higher than in turbulent conditions when $s>0$, activation in the turbulent case is more efficient over the range of $\overline{s}$. This enhancement in activation from small-scale turbulent fluctuations suggests that cloud-scale LES models, as well as global-scale parameterizations based on these LES models, will underestimate the number of aerosol particles forming CCN.

\subsection*{Increased CCN Concentrations due to Evaporation Dynamics}
If the dynamics of evaporation are neglected and, instead, droplets are assumed to reach equilibrium with the environment instantaneously, particles in regions with $s\ge s_{\text{crit}}$ would be CCN-active, whereas particles in regions with $s<s_{\text{crit}}$ would be CCN-inactive. This was the assumption applied in \cite{shawon2021dependence,prabhakaran2020role}. However, in a real cloud, droplets do not respond instantaneously to fluctuations in supersaturation, which leads to further enhancements in CCN activation (see comparison between dashed and solid green lines in Fig.~\ref{fig:activated_fraction_mean_S}c). When a droplet enters a region of subsaturated air, the particle will evaporate according to a characteristic timescale $\tau_{\text{evap}}=-\overline{D_{\text{p}}/2}^2/(2\xi_1 s)$, where $\xi_1$ is the normalized condensation growth parameter \cite{yau1996short}. If $\tau_{\text{evap}}=0$, droplets are assumed to instantly respond to perturbations in $s$.

Across most $\overline{s}$, we find the activated fraction is greater in our simulations (solid green line) than it would be if droplets were assumed to respond to supersaturation fluctuations instantaneously (i.e., $\tau_{\text{evap}}=0$; dashed green line). The deviation between these two cases depends on the magnitude of $\tau_{\text{evap}}$ relative to the autocorrelation timescale of the supersaturation experienced by the droplets, $\tau_{\text{s}}$. Using $s$ along the ensemble of Lagrangian parcels, $\tau_{\text{s}}\approx7.5$~s, whereas $\tau_{\text{evap}}=9.9$~s and $\tau_{\text{evap}}=120$~s for $\overline{s}=-1.2\%$ and $\overline{s}=-0.2\%$, respectively. Because $\tau_{\text{evap}}$ is longer than $\tau_{\text{s}}$, particles remain CCN-activated for a significant period of time before evaporating, consistent with the findings in Siewert et al. \cite{siewert2017statistical}. This delay in evaporation leads to further increases in the concentration of CCN within the turbulent environment.

\section*{Discussions and Conclusions}

The similarity in the shapes of the solid and dashed green lines in Fig. \ref{fig:activated_fraction_mean_S}c shows our results include the three activation regimes described in Prabhakaran et al. \cite{prabhakaran2020role}. Because droplets in our simulation do not immediately respond to fluctuations, we find these regimes are shifted towards lower $s$ than they report. The presence of these regimes implies that different regions of a cloud are more likely to be impacted by fluctuations in $s$ than others. For example, Grabowski et al. \cite{grabowski2022impact} recently studied activation of CCN in an updraft by using adiabatic parcels that include turbulent variability in $s$. They concluded that fluctuations in $s$ only activate a fraction of CCN. However, this conclusion may be a consequence of $\overline{s}$ increasing in the updraft to the point where the magnitude of $\overline{s}$ overwhelms the effect of fluctuations. Our results agree with the finding that fluctuations may not be important if cloud updrafts and, thereby $\overline{s}$, are high enough that nearly all particles activate. Instead the fluctuations may be more relevant along cloud edges or under weak updrafts, where $\overline{s}$ is lower.

In conclusion, the enhancements in CCN activation shown here may have a large impact on the number concentration of CCN simulated by atmospheric models, which have long been known to affect precipitation rates and the radiative properties of clouds \cite{twomey1977influence,albrecht1989aerosols, rosenfeld2008flood}. Earth System Models and cloud-scale LES do not resolve the isobaric mixing of parcels with different temperatures and water vapor concentrations and, thereby, neglect the small-scale variability in $s$. We show that these simplifications may lead to misrepresentation of modeled CCN concentrations, which may contribute to the large uncertainties in predictions of aerosol-cloud interactions.

\matmethods{

\subsection{LES of Pi Chamber}

We use LES output from the System for Atmospheric Modeling (SAM) \cite{khairoutdinov2003cloud} to calculate parcel trajectories. SAM was previously adapted for the Pi Chamber, which is described in \cite{Thomas2019,yang2022large}. The LES domain is a 2~x~2~x~1~m$^3$, with a grid spacing of 3.125~cm and time step of 0.02~s. The top and bottom surface temperatures are set as 280 and 299 K, and both surface are saturated with water vapor. The side wall is set to 285 K with a wall relative humidity of 78\% such that the mean supersaturation without cloud droplets is approximately 2.5\%, consistent with chamber observations. In both the chamber and the LES, monodisperse salt particles are injected into the chamber at a constant rate. A spectral-bin microphysics scheme is used to represent the droplet size distribution and its interaction with the environment, which is critical to determining the fluctuating water vapor and temperature fields with the chamber. Notably, the particles in the LES are separate from our box model simulations and are not included in our analysis. However, droplet microphysics must be included in the LES to ensure the water vapor and temperature fields include the effects of condensation onto particles and the evaporation of droplets. The simulated cloud and the dynamic and thermodynamic fields reach a steady state after about 5 min. The total simulation is 1 hour and the 3D velocity and scalar fields are output every 25th time step (i.e., every 0.5 seconds). These scalars are used to calculate the motion and growth of particles in each Lagrangian trajectory.

\subsection*{Parcel Trajectories from LES}
Velocity fields from the LES were used to compute Lagrangian trajectories within the chamber. To quantify the impact of the turbulent environment on the particle populations over time, we simulated the simultaneous release of 10,000 Lagrangian tracers at random locations within the chamber and tracked the evolution of the environmental properties within the parcel. The temporal evolution of the parcel location is computed by solving the equations of motion for the parcel:
\begin{equation}
\frac{dx_{\text{p}}(t)}{dt}=u_{\text{f}}(t)
\qquad
\frac{dy_{\text{p}}(t)}{dt}=v_{\text{f}}(t)
\qquad
\frac{dz_{\text{p}}(t)}{dt}=w_{\text{f}}(t),
\label{eqn:dxdt}
\end{equation}


\noindent where $x_{\text{p}}(t)$, $y_{\text{p}}(t)$, and $z_{\text{p}}(t)$ are the three-dimensional coordinates of the parcel over time and $u_{\text{f}}(t)$, $u_{\text{f}}(t)$, and $w_{\text{f}}(t)$ are the $x$, $y$, and $z$ components of fluid velocity experienced by the parcel over time. In this case, inertial effects are neglected, and the air velocity simulated by the LES was assumed to be the same as the particle velocity. Though we neglected gravitational settling when calculating parcel trajectories, we can represent the decay in number concentration associated with each Lagrangian parcel through Stoke's settling. However, to facilitate comparison with the uniform supersaturation approximation, we did not include particle and droplet removal when calculating the fraction activated.

A linear interpolator was used to compute $u(t)$, $u(t)$, and $w(t)$ from the 4D LES velocity fields, given the simulation time, $t$, and parcel location, $x_{\text{p}}(t)$, $y_{\text{p}}(t)$, and $z_{\text{p}}(t)$. The temporal evolution of each parcel's position was calculated using the initial value ordinary differential equation solver in the SciPy library  with a maximum time step of 0.1~s. As the parcel was transported through the chamber, a linear interpolator was used to compute the temporal evolution of the fluid temperature $T(t)$ and water vapor mixing ratio $q(t)$ experienced by the parcel. The water vapor supersaturation within the parcel, $s(t)$ is given by:

\begin{equation}
\label{s}
    s(t) = \frac{e(t)}{e_{\text{sat}}(t)}-1,
\end{equation}

\noindent where the water vapor pressure, $e(t)$, along the trajectory is given by: 

\begin{equation}
    e(t)=p\frac{q}{0.62q},
\end{equation}

\noindent and the saturation water vapor pressure ($e_{\text{sat}}$) is given by:
\begin{equation}
\label{eqn:e_sat}
    e_{\text{sat}} = 0.61094 \exp\left(\frac{17.625 (T-273.15)}{(T-30.11)}\right).
\end{equation}

\noindent The saturation pressure and atmospheric pressure $p$ are expressed in kPa. Thermodynamic relationships were adapted from the python-based cloud parcel model, pyrcel \cite{rothenberg2016metamodeling}.

\subsection*{Simulation of droplet activation and growth}
The temporal evolution of droplet diameter, $D_p$, for an individual particle is modeled using a subset of equations for an adiabatic parcel model \cite{nenes2001kinetic,seinfeld2008atmospheric}, as implemented in Rothenberg et al.\cite{rothenberg2016metamodeling}. In this case, we use the environmental fields from the LES to drive the parcel simulations. The particle growth/evaporation rate is given by:
\begin{equation}
    \label{eqn:drdt}
    \frac{dD_p}{dt}=\frac{G}{D_p}(s-s_{\text{eq}}),
\end{equation}
\noindent where $dD_i/dt$ is positive and the droplet grows when the environmental supersaturation, $s$, is greater than the equilibrium supersaturation at the droplet surface, $s_{\text{eq}}$. The growth coefficient $G$ is given by:
\begin{equation}
    \label{eqn:G}
        G=\left(\frac{\rho_{\text{w}}RT}{e_{\text{sat}}D'_vM_{\text{w}}}+\frac{L\rho_{\text{w}}[(LM_{\text{w}}/RT)-1]}{k'_aT}\right)^{-1},
\end{equation}
\noindent where $M_{\text{w}}$ is the molecular weight of water, $R$ is the universal gas constant, and $e_{\text{sat}}$ is given in Eqn.~\ref{eqn:e_sat}.  \cite{rothenberg2016metamodeling}, the diffusivity and conductivity of air, denoted $D_v$ and $k_a$, respectively, are modified to account for non-continuum effects. The modified expressions are given by:
\begin{equation}
    \label{eqn:Dv}
    D'_v=D_v\left(1+\frac{D_v}{a_cr}\sqrt{\frac{2\pi M_{\text{w}}}{RT}}\right)^{-1}
\end{equation}

\noindent and

\begin{equation}
    k'_a=k_a\left(1+\frac{k_a}{a_Tr\rho_{\text{a}}c_p}\sqrt{\frac{2\pi M_{\text{w}}}{RT}}\right)^{-1},
\end{equation}
where $a_T$ is  the thermal accommodation coefficient and $a_c$ is the condensation coefficient, which are assumed to be 0.96 and 1, respectively.

The equilibrium supersaturation of an aqueous droplet, $s_{\text{eq}}$, is computed given by:
\begin{equation}
    s_{\text{eq}}=\frac{D_p^3-D_d^3}{D_p^3-D_d^3(1-\kappa)}\exp\left( \frac{4\sigma_{s/a} M_w}{R T_p \rho_w D_p} \right)-1,
\end{equation}

\noindent where $\sigma_{s/a}$ is the surface tension of the particle,  $\rho_w$ is the density of water, $D_{\text{dry}}$ is the dry aerosol diameter, and $\kappa$ is the effective hygroscopicity parameter of the aerosol mixture. We assumed $\kappa=1$ for the NaCl particles, which each had $D_{\text{dry}}=130~nm$. The equations were solved using the initial value problem solver in the SciPy library.

}

\showmatmethods{} 

\acknow{L. Fierce and M. Ovchinnikov were supported by the U.S. Department of Energy’s Atmospheric System Research, an Office of Science Biological and Environmental Research program. The Pacific Northwest National Laboratory (PNNL) is operated for DOE by Battelle Memorial Institute under contract DE-AC05-76RLO1830. J. Anderson and P. Beeler were supported by the U.S. Department of Energy, Office of Science, Office of Workforce Development for Teachers and Scientists, Office of Science Graduate Student Research (SCGSR) program. The SCGSR program is administered by the Oak Ridge Institute for Science and Education (ORISE) for the DOE. ORISE is managed by ORAU under contract number DE‐SC0014664. All opinions expressed in this paper are the author’s and do not necessarily reflect the policies and views of DOE, ORAU, or ORISE. The MTU coauthors were supported by U.S. National Science Foundation grants AGS-1754244 and AGS-2133229. S.\ Krueger was supported by the National Science Foundation (managed by Michigan Technological University) under Grant AGS-2133229. F. Yang was supported by the Office of Biological and Environmental Research in the Department of Energy, Office of Science, through the United States Department of Energy Contract No. DE-SC0012704 to Brookhaven National Laboratory.}

\showacknow{} 

\bibliography{pnas-sample}

\end{document}